\documentclass[aps,pre,twocolumn,showpacs]{revtex4}
\usepackage{epsfig}
\usepackage{times}
\begin{document}

\title{Dynamical mean-field approximations for a diffusive pair contact process}

\author{Attila Szolnoki}
\affiliation
{Research Institute for Technical Physics and Materials Science
P.O. Box 49, H-1525 Budapest, Hungary}


\begin{abstract}
Dynamical mean-field approximations are performed to study the 
phase transition of a pair contact process with diffusion in different
spatial dimensions. The level of approximation is extended up to $18$-site 
clusters for the one-dimensional model.
The application of coherent anomaly method 
shows that the critical $\beta$ exponent does not depend on
the strength of diffusion rate. The extension of the mean-field approximation
to higher dimensions also suggests that the critical behavior 
may be described by a unique set of exponents.
\end{abstract}

\pacs{05.70.Ln, 64.60.Ak}

\maketitle

The study of nonequilibrium critical phenomena has received considerable
attention but their understanding is far from complete.
Even the simplest models can be theoretically challenging.
A well-known example is 
the pair contact process model with diffusion (PCPD) \cite{carlon:pre01}.
Despite its intensive study there is no satisfying consensus about 
the accurate values of critical exponents
(see Ref.~\cite{henkel:jpa04}, for a recent review and further references). 
Different methods and approximations were applied, such as 
Monte Carlo simulation (MC) 
\cite{hinrichsen:pre01,henkel:jpa01,park:pre02,odor:pre00}, 
dynamical mean-field approximation (DMF) \cite{odor:pre00,carlon:pre01}, 
density-matrix renormalization group study \cite{carlon:pre01}, 
and field-theoretic approach \cite{wijland:cm03,janssen:cm04}.
Most of the studies have been devoted to the one-dimensional case but
the published numerical results are partially contradictory.
In addition to the simulation difficulties, a consistent field theory
description of the model is still an unresolved issue \cite{janssen:cm04}.
Instead of summing all the raised proposals \cite{henkel:jpa04}, we consider
one of the suggestions that the value of critical exponent may depend on 
the strength of diffusion. This idea was inspired by different MC simulations 
\cite{dickman:pre02,odor:pre00} and DMF approximations \cite{odor:pre00}.
This latter approach, however, left some ambiguities   
for strong or very weak diffusion rate \cite{odor:pre00}.
Subsequent simulations \cite{kockelkoren:prl03,odor:pre03} 
revealed the concept of a novel universality class with a unique set of exponents. 
However exact calculation, albeit for a slightly different model, suggested 
the possibility of two transitions depending on the diffusion rate
in higher spatial dimensions \cite{paessens:jpa04}. 
The primary purpose of this Brief Report to clarify the mentioned ambiguities
of DMF approach by significantly improving the level of one-dimensional 
approximations and to extend the calculations to higher spatial dimensions.
 
The simplicity of PCPD model, especially in one dimension, makes possible 
to extend the DMF approximation to large-size clusters. 
Such a large number of approximations for the order parameter function 
enables the application of the coherent anomaly method (CAM) to derive
quantitative prediction to the value of $\beta$ exponent \cite{suzuki:jpj86}. 
Another challenge is to develop the DMF approximation to higher spatial
dimensions. 
Generally, such kind of calculation
is expected to yield qualitatively correct results in high dimensions, 
therefore it can help us to clarify the raised question whether the values of 
critical exponents depend on the strength of diffusion rate.

The PCPD model \cite{carlon:pre01,howard:jpa97} is the extension of 
the pair-contact process model (PCP) \cite{jensen:prl93} that belongs to 
the well-known DP universality class \cite{munoz:prl96}.
The PCPD model consists of a $d$-dimensional lattice, with periodic boundaries,
where each site is either vacant or occupied by a single particle. 
The dynamical rules for updating the system are defined as follows.
A randomly chosen pair of nearest-neighbor particles is annihilated with a 
probability $p(1-D)$. An additional particle is created around the given
pair with probability $(1-p)(1-D)$ if it is not forbidden by the
exclusion principle. Furthermore individual particles are allowed to
hop to the nearest-neighbor empty site with diffusion rate $D$.
This version is also called restricted PCPD model because the exclusion 
law prohibits double occupancy.
In this model the order parameter is 
the concentration of pairs of particles ($u$) that becomes zero above a
critical value of $p$ when the system arrives to one of the two possible
absorbing states \cite{henkel:jpa04}.

\begin{figure}
\centerline{\epsfig{file=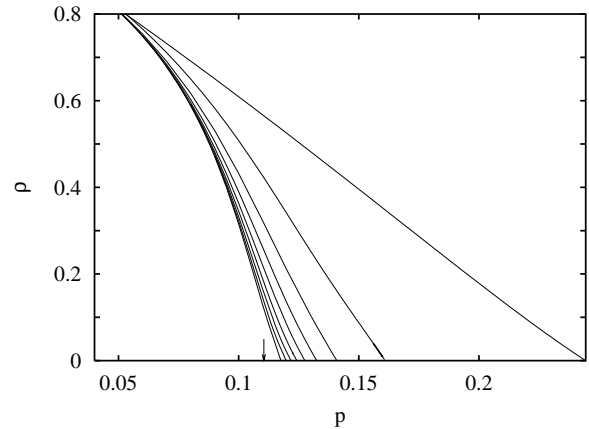,width=8.5cm}}
\caption{\label{fig:mfig1}Dynamical mean-field approximations of the 
particle concentration at $D=0.1$ for $n=2,4,6, ..., 18$-point levels 
(from right to left curves). The arrow shows the location of the
extrapolated phase transition point.}
\end{figure}

The application of DMF approximation, which is a dynamical version of the
cluster variation method, involves finding a hierarchy of evolution equations
for the probability distributions of configurations within a cluster of
$n$ sites (for details, see e.g. Refs.~\cite{dickman:pla87,marro:99}).
The calculation of transitions of cluster probabilities would generate
infinite hierarchy of equations, since transitions in an $n$-site cluster
may depend on sites outside the cluster. To avoid this the $n$-site 
approximation estimates $m$-site probabilities (for $m > n$) by using Bayesian
extension process.
This type of approximation has proved to be successful in the evaluation 
of phase diagram for many different nonequilibrium models
\cite{gutowitz:pd87,szabo:pra91,benavraham:pra92,szabo:pre94,marques:pre02,szolnoki:pre00,szolnoki:pre02,szabo:pre98}. 
Evidently, this method was also applied to the present model in one dimension
up to $n = 7$, however, the corresponding results were insufficient 
to derive reliable prediction to the values of critical exponents 
\cite{carlon:pre01,odor:pre03}. Furthermore
CAM result could not be presented for strong ($D>0.5$) and for weak ($D<0.1$)
diffusion rates \cite{odor:pre00}. 

\begin{widetext}
\begin{center}
\begin{table}
\caption{\sf Results of $n$-site approximations for the one-dimensional model. 
The extrapolated values of $p_c$ and the order parameter exponents $\beta$
for different values of diffusion rate ($D$) are also given.
Numbers in parentheses correspond to the uncertainty in the last digit.}
\begin{tabular}{lllllllll} 
\hline
\hline
$n \backslash D$ \,& \hspace{1.0cm} $0.05$ & & \hspace{1.2cm} $0.1$ & & \hspace{1.2cm} $0.5$ & & \hspace{1.2cm} $0.85$ & \\
 & \hspace{0.5cm} $p_c^n$ & \hspace{0.2cm} $a_n$ & \hspace{0.5cm} $p_c^n$ & \hspace{0.2cm} $a_n$ & \hspace{0.5cm} $p_c^n$ & \hspace{0.2cm} $a_n$ & \hspace{0.5cm} $p
_c^n$ & \hspace{0.2cm} $a_n$ \\
\hline
4 & 0.157167(1) &\,\, 48.4(1) &\,\,\, 0.160964(1) &\,\, 82.1(2) &\,\,\, 0.209694(2) &\,\, 51.7(1) &\,\,\, 0.294937(1) &\, 17.61(1) \\
6 & 0.137271(1) &\,\, 94.2(4) &\,\,\, 0.140885(1) & 147.3(6) &\,\,\, 0.184175(4) &\,\, 93.2(2) &\,\,\, 0.274965(2) &\, 17.40(2) \\
8 & 0.129264(2) & 127.1(6) &\,\,\, 0.132484(4) & 207.6(10) &\,\,\, 0.171820(6) & 139.2(8) &\,\,\, 0.260674(6) &\, 19.63(2) \\
10 & 0.124488(5) & 158.2(10) &\,\,\, 0.127542(4) & 265.5(18) &\,\,\, 0.164401(6) & 190.9(10) &\,\,\, 0.250095(8) &\, 23.08(3) \\
12 & 0.121299(7) & 205.9(20) &\,\,\, 0.124240(10) & 330.8(16) &\,\,\, 0.159422(8) & 248.9(14) &\,\,\, 0.242122(12) &\, 26.83(4) \\
14 & 0.119015(7) & 254.8(20) &\,\,\, 0.121799(10) & 413.6(30) &\,\,\, 0.155843(10) & 308.0(20) &\,\,\, 0.235710(16) &\, 33.31(10) \\
16 & 0.11730(3) & 308.2(60) &\,\,\, 0.11996(3) & 489.4(60) &\,\,\, 0.15309(4) & 367.2(40) &\,\,\, 0.23071(3) &\, 41.02(20) \\
18 & 0.11596(4) & 370.5(50) &\,\,\, 0.11841(11) & 590.2(80) &\,\,\, 0.15053(10) & 453.3(70) &\,\,\, 0.22676(12) &\, 49.19(40) \\
\hline
$p_c$ & 0.1053(2) &  &\,\,\, 0.1066(5) &  &\,\,\, 0.1335(1) &  &\,\,\, 0.1955(5) &  \\
$\beta$ &  & 0.54(4) &  & 0.55(6) &  & 0.51(4) &  &\,\, 0.48(3) \\
\hline
\hline
\end{tabular}
\label{table}
\end{table}
\end{center}
\end{widetext}

To eliminate these shortages now the approximation is extended up to $n=18$ 
level. Naturally, for higher level approximations the equations can 
be only build and solved numerically \cite{dickman:pre02b,szolnoki:pre02}. 
The stationary distribution is attained
via numerical integration when the sum of the absolute values of all time
derivatives smaller than $\epsilon$ (generally $\epsilon = 10^{-8}$ is used).
Increasing the level of approximation the necessary number of iterations 
to attain the steady-state configurations probabilities increases drastically.
To demonstrate the speed of convergence for large $n$, four-week running is 
necessary to get data at $18$-point level.
Despite the time consuming convergence during the iterations,
the calculations are tractable even by a personal computer. 
Figure~\ref{fig:mfig1} illustrates the improvement of solutions
as we increase the levels of approximation.   
Here the particle concentration functions $\rho(p)$ are plotted 
for $D=0.1$, which show linear decay around the transition point.
The concentration of pairs can be displayed in a similar figure but
$u(p)$ decreases quadratically around $p_c^n$ for $n>2$.

The order-parameter exponent can be estimated from the mean-field data
following the CAM approach developed by Suzuki \cite{suzuki:jpj86}. 
In the vicinity of
the critical point the order parameter function can be fitted as
$u(p) \approx a_n\,(p_c^n-p)^{\beta^{MF}}$, where $p_c^n$ denotes the 
value of the critical point at $n$-point level and $p_c$ is the
extrapolated value to the $n \to \infty$ limit. As we mentioned 
$\beta^{MF}=2$ is valid for all values of $D$ if $n>2$.
Table~\ref{table} summarizes the results for the estimated transition
points ($p_c^n$) and amplitudes ($a_n$) at different levels of 
approximation for various values of diffusion rate. (The starting point
of fitting is determined when the corresponding particle concentration 
drops below 0.01.)
According to the CAM analysis the amplitudes $a_n$ scales as
\begin{equation} 
a_n \approx (p_c^n - p_c)^{\beta - \beta^{MF}} .
\label{eq:cam}
\end{equation} 
\begin{figure}
\centerline{\epsfig{file=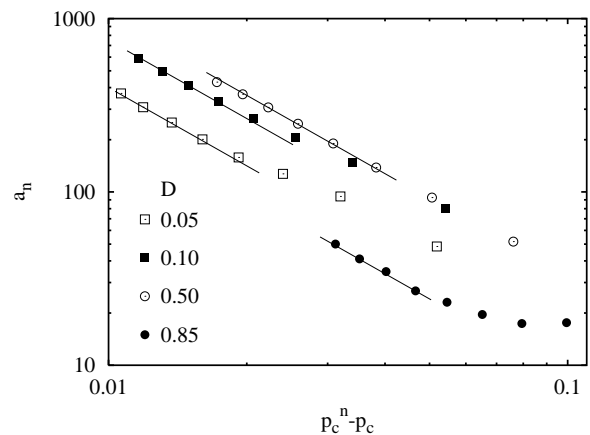,width=8.5cm}}
\caption{\label{fig:mfig2}CAM scaling of the critical mean-field coefficients
for the order parameter at different values of the diffusion constant. 
The lines correspond to the $\beta = 0.5$ exponent. The results of 
$n=4,6,8,...,18$-point level of approximations are plotted from right to
left.}
\end{figure}

As a consequence, the $log - log$ plot of the amplitudes {\it vs.}
$p_c^n - p_c$ yields an estimation for the $\beta$ exponent. 
Figure~\ref{fig:mfig2} shows these plots for different values of $D$.
This figure indicates clearly that wrong value of $\beta$ could be deduced
by this technique if the calculations are limited to small-size clusters.
For large $n$, however, the power law fits of the corresponding
amplitudes become possible. 
The deviation of $a_n$ from an expected power law function is 
pronounced for large $D$ when $n$ are small.
Although the convergence of $a_n$ is obvious, still the uncertainty of the 
estimated $\beta$ value is close to $10 \%$. 
The extrapolated values of $p_c$ and the $\beta$ exponents for different 
diffusion rates are also given in Table~\ref{table}. 
Within the mentioned uncertainty the same $\beta=0.50(5)$ exponents
can be fitted for all values of $D$ as displayed in Fig.~\ref{fig:mfig2}.
This $\beta$ value is comparable 
to some earlier published MC results \cite{odor:pre00,dickman:pre02}. 
To end this section it is worth emphasizing the sensitivity of CAM on
the levels of approximation. 
The comparison of Fig.~\ref{fig:mfig1} and Fig.~\ref{fig:mfig2} underlines
the different behaviors of $p_c^n$ and $a_n$ as a function of $n$.
As an example, the approach of $a_n$ to the power law function is significant
if we increase $n$ from $8$ to $18$ for $D=0.1$ but
the location of the transition point changes less then $11 \%$.  
Similar behavior was reported by a recent study \cite{park:cm04} where the
same model was investigated by DMF approximation up to $n=13$ for $D=0.5$.

In the light of a field-theoretic calculation \cite{paessens:jpa04} it is also
interesting to extend our investigations to higher spatial dimensions ($d>1$). 
The main question is whether two transitions exist as a function of 
diffusion rate, $D$. 
The lowest level that considers the value of $d$ is the pair approximation 
($n=2$).
In spite of its failure in one dimension \cite{carlon:pre01,henkel:jpa04}, 
this approximation is expected to provide qualitatively correct
behavior in higher dimensions. At this level the two
independent variables are the particle concentration denoted by $\rho$ and
the concentration of pairs $u$. 
The equations of their time evolution on a $d$ dimensional lattice 
can be written as
\begin{widetext}
\begin{eqnarray} 
\dot{\rho} &=& -2(1-D)p\,u +(1-D)(1-p) \frac{u(\rho-u)}{\rho} \\
\dot{u} &=& -(1-D)p\,\frac{u}{(\rho)^{2m}} 
\sum_{k=0}^{2m}(k+1){2m\choose k}(u)^k(\rho-u)^{2m-k} \nonumber \\ 
& &+(1-D)(1-p)\,\frac{u(\rho-u)}{\rho(1-\rho)^m} \sum_{k=0}^m (k+1) {m\choose k} 
 (\rho -u)^k (1-2\rho + u)^{m-k} \nonumber \\
& &+D\,\frac{\rho-u}{\rho^m (1-\rho)^m}
 \sum_{f=0}^m {m\choose f} (\rho-u)^f (1-2\rho+u)^{m-f} \left( \sum_{b=0}^m (f-b) {m\choose b} u^b (\rho-u)^{m-b}\right)\,\,,
\label{eq:pair}
\end{eqnarray}  
\end{widetext}
where $m=2d-1$.
By comparing these equations in case of $d=1$ to the equations in 
Ref.~\cite{henkel:jpa04},
one can observe that a prefactor $2$ is missing from the diffusion
part of Eq.~\ref{eq:pair}.
The slight difference can be explained as follows.
When doing MC simulation an elementary diffusion process includes
to choose a particle {\it and} a direction as well.
Afterwards the jump is executed
with probability $D$ if the target site is empty. Therefore,
the probability to jump to a specified direction is equal to $D/(2d)$. 
The normalization is fulfilled by choosing the latter jumping rate.

Returning now to the above equations the solutions in the active phase
are
\begin{eqnarray} 
\rho &=& \frac{1-p}{1-3p}\,\, u \,\,\,, \,\,\mbox{\rm where} \nonumber\\
\\
u &=& \left(\frac{1-3p}{1-p}\right)^2 
\frac{2m - (1-D) (2m\,3p+1-p)}{\,\,2m- (1-D) (2m\,3p+1-3p)} \nonumber 
\end{eqnarray}  
As a consequence, the value of critical point is
\begin{equation} 
p_{c}^{MF}(d,D) = \left\{ \begin{array}{cl} 
\frac{4d-3+D}{(12d-7)(1-D)}\,, &\mbox{\rm if}\,\,\,0\leq D \leq \frac{1}{2(3d-1)}  \\ 
\\ 
\frac{1}{3}\,, &\mbox{\rm otherwise}\,. 
\end{array} \right. 
\end{equation} 
It means that, as in one dimension, two different regions can be detected at 
$any$ finite values of $d$ as a function of $D$. 
Nevertheless, the region that characterizes the high-$D$ behavior,
including $p_c=1/3$, expands if we increase $d$ and in the infinite
dimension limit the solution convergences to the solution of the one-site 
approximation. 
From this result one may conjecture that there is no different universality
classes at low and high diffusion rate for the restricted PCPD in higher
dimensions.
\begin{figure}
\centerline{\epsfig{file=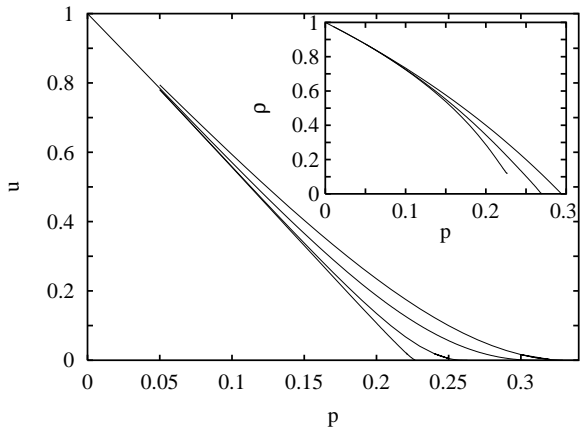,width=8.5cm}}
\caption{\label{fig:mfig3}The pair-density function of the two-dimensional
model at $9$-point level for $D=0, 0.1, 0.5$ and $0.9$ values 
(from left to right curves).
The inset shows particle density  at $D=0$ for $n=2, 4$, and $9$-point 
levels (from right to left curves).}
\end{figure}

More convincing arguments can be achieved by performing DMF 
approximation for larger clusters in two dimensions. 
In the absence of diffusion ($D=0$) the two-point
approximation yields qualitatively incorrect result, since the particle 
concentration
becomes zero at the transition point ($p_c^{2p}=5/17 \approx 0.2941)$.
This is contrary to the MC result that revealed a finite
'natural' density, $\rho_{nat}=0.1477$
at the transition point ($p_c=0.2005$) for the PCP model \cite{kamphorst:pre99}. 
The four-site approximation (on $2 \times 2$ clusters) yields a better estimation
for the transition point ($p_c^{4p}=0.2691(3)$), however, it also predicts
zero particle concentration at the transition point. 
This latter failure is already diminished at nine-site
($3 \times 3$ clusters) approximation. As the inset of
Fig.~\ref{fig:mfig3} shows, this level of approximation can already describe
qualitatively well the behavior of the two-dimensional PCP model:
it gives $\rho_{nat}=0.1197(2)$ at $p_c^{9p}=0.22607(3)$.  
The reason why the $9$-point level gives a qualitatively different
result may be explained by the fact that
the $3 \times 3$ cluster size is the smallest cluster that covers all 
the elementary processes. (For example, a particle creation at the ends 
of an occupied pair "requires" a minimum size of $3$ in one direction.) 
Similar qualitative improvement of the results have already been observed 
in a two-offspring branching annihilating random walk model 
\cite{zhong:epjb03}. In the latter one-dimensional model at least
$5$-point level of approximation was necessary for the correct description.
The success of $9$-point level of approximation also means that the 
smallest cluster that may give correct description 
of our PCPD model in three dimensions consists $3 \times 3 \times 3 = 27$ 
lattice points. This level is not obtainable presently.
Returning to the two-dimensional case, the $9$-point approximation
does not predict qualitatively different behavior for low and high 
diffusion rate. Figure~\ref{fig:mfig3} demonstrates that 
the value of critical point increases with $D$ 
($p_c^{9p}(D=0.1)=0.2607(5), \, p_c^{9p}(D=0.9)=0.3315(5)$)
while the order parameter function does not differ qualitatively for 
$D=0.1$ and $D=0.9$. 
This result also supports the single universality class conjecture. 

To summarize, DMF approximation is extended to large-cluster levels 
to study the critical behavior of the one-dimensional restricted PCPD model.
We find that the ambiguity of CAM approach for weak and strong $D$ 
can be resolved by increasing the level of approximation.
The approximation suggests uniform $\beta$ exponent for all values of $D$
within the uncertainty of the estimations.
This $\beta = 0.50(5)$ value is comparable with some published MC 
results and supports that the model leaves the DP universality class and 
described by an exponent that also differs from the PC class. 
The approximations in higher dimensions also
support the conjecture that the PCPD model can be described by 
a single universality class that is independently from the rate of diffusion.

\begin{acknowledgments}
I thank G\'eza \'Odor (who suggested this investigation), and Gy\"orgy
Szab\'o for illuminating discussions.
This work was supported by the Hungarian National Research Fund under 
Grant No. T-47003.
\end{acknowledgments}


\begin{thebibliography}{30}
\expandafter\ifx\csname natexlab\endcsname\relax\def\natexlab#1{#1}\fi
\expandafter\ifx\csname bibnamefont\endcsname\relax
  \def\bibnamefont#1{#1}\fi
\expandafter\ifx\csname bibfnamefont\endcsname\relax
  \def\bibfnamefont#1{#1}\fi
\expandafter\ifx\csname citenamefont\endcsname\relax
  \def\citenamefont#1{#1}\fi
\expandafter\ifx\csname url\endcsname\relax
  \def\url#1{\texttt{#1}}\fi
\expandafter\ifx\csname urlprefix\endcsname\relax\def\urlprefix{URL }\fi
\providecommand{\bibinfo}[2]{#2}
\providecommand{\eprint}[2][]{\url{#2}}

\bibitem[{\citenamefont{Carlon et~al.}(2001)\citenamefont{Carlon, Henkel, and
  Schollw\"ock}}]{carlon:pre01}
\bibinfo{author}{\bibfnamefont{E.}~\bibnamefont{Carlon}},
  \bibinfo{author}{\bibfnamefont{M.}~\bibnamefont{Henkel}}, \bibnamefont{and}
  \bibinfo{author}{\bibfnamefont{U.}~\bibnamefont{Schollw\"ock}},
  \bibinfo{journal}{Phys. Rev. E} \textbf{\bibinfo{volume}{63}},
  \bibinfo{pages}{036101} (\bibinfo{year}{2001}).

\bibitem[{\citenamefont{Henkel and Hinrichsen}(2004)}]{henkel:jpa04}
\bibinfo{author}{\bibfnamefont{M.}~\bibnamefont{Henkel}} \bibnamefont{and}
  \bibinfo{author}{\bibfnamefont{H.}~\bibnamefont{Hinrichsen}},
  \bibinfo{journal}{J. Phys. A} \textbf{\bibinfo{volume}{37}},
  \bibinfo{pages}{R117} (\bibinfo{year}{2004}).

\bibitem[{\citenamefont{Hinrichsen}(2001)}]{hinrichsen:pre01}
\bibinfo{author}{\bibfnamefont{H.}~\bibnamefont{Hinrichsen}},
  \bibinfo{journal}{Phys. Rev. E} \textbf{\bibinfo{volume}{63}},
  \bibinfo{pages}{036102} (\bibinfo{year}{2001}).

\bibitem[{\citenamefont{Henkel and Schollw\"ock}(2001)}]{henkel:jpa01}
\bibinfo{author}{\bibfnamefont{M.}~\bibnamefont{Henkel}} \bibnamefont{and}
  \bibinfo{author}{\bibfnamefont{U.}~\bibnamefont{Schollw\"ock}},
  \bibinfo{journal}{J. Phys. A} \textbf{\bibinfo{volume}{34}},
  \bibinfo{pages}{3333} (\bibinfo{year}{2001}).

\bibitem[{\citenamefont{Park and Kim}(2002)}]{park:pre02}
\bibinfo{author}{\bibfnamefont{K.}~\bibnamefont{Park}} \bibnamefont{and}
  \bibinfo{author}{\bibfnamefont{I.-M.} \bibnamefont{Kim}},
  \bibinfo{journal}{Phys. Rev. E} \textbf{\bibinfo{volume}{66}},
  \bibinfo{pages}{027106} (\bibinfo{year}{2002}).

\bibitem[{\citenamefont{\'Odor}(2000)}]{odor:pre00}
\bibinfo{author}{\bibfnamefont{G.}~\bibnamefont{\'Odor}},
  \bibinfo{journal}{Phys. Rev. E} \textbf{\bibinfo{volume}{62}},
  \bibinfo{pages}{R3027} (\bibinfo{year}{2000}).

\bibitem[{\citenamefont{van Wijland et~al.}(2003)\citenamefont{van Wijland,
  T{\"a}uber, and Deloubri{\'e}re}}]{wijland:cm03}
\bibinfo{author}{\bibfnamefont{F.}~\bibnamefont{van Wijland}},
  \bibinfo{author}{\bibfnamefont{U.~C.} \bibnamefont{T{\"a}uber}},
  \bibnamefont{and}
  \bibinfo{author}{\bibfnamefont{O.}~\bibnamefont{Deloubri{\'e}re}},
  \bibinfo{journal}{condmat/0311568}  (\bibinfo{year}{2003}).

\bibitem[{\citenamefont{Janssen et~al.}(2004)\citenamefont{Janssen, van
  Wijland, Deloubri{\'e}re, and T{\"a}uber}}]{janssen:cm04}
\bibinfo{author}{\bibfnamefont{H.-K.} \bibnamefont{Janssen}},
  \bibinfo{author}{\bibfnamefont{F.}~\bibnamefont{van Wijland}},
  \bibinfo{author}{\bibfnamefont{O.}~\bibnamefont{Deloubri{\'e}re}},
  \bibnamefont{and} \bibinfo{author}{\bibfnamefont{U.~C.}
  \bibnamefont{T{\"a}uber}}, \bibinfo{journal}{condmat/0408064}
  (\bibinfo{year}{2004}).

\bibitem[{\citenamefont{Dickman and de~Menezes}(2002)}]{dickman:pre02}
\bibinfo{author}{\bibfnamefont{R.}~\bibnamefont{Dickman}} \bibnamefont{and}
  \bibinfo{author}{\bibfnamefont{M.~A.~F.} \bibnamefont{de~Menezes}},
  \bibinfo{journal}{Phys. Rev. E} \textbf{\bibinfo{volume}{66}},
  \bibinfo{pages}{045101} (\bibinfo{year}{2002}).

\bibitem[{\citenamefont{Kockelkoren and Chat\'e}(2003)}]{kockelkoren:prl03}
\bibinfo{author}{\bibfnamefont{J.}~\bibnamefont{Kockelkoren}} \bibnamefont{and}
  \bibinfo{author}{\bibfnamefont{H.}~\bibnamefont{Chat\'e}},
  \bibinfo{journal}{Phys. Rev. Lett.} \textbf{\bibinfo{volume}{90}},
  \bibinfo{pages}{125701} (\bibinfo{year}{2003}).

\bibitem[{\citenamefont{\'Odor}(2003)}]{odor:pre03}
\bibinfo{author}{\bibfnamefont{G.}~\bibnamefont{\'Odor}},
  \bibinfo{journal}{Phys. Rev. E} \textbf{\bibinfo{volume}{67}},
  \bibinfo{pages}{016111} (\bibinfo{year}{2003}).

\bibitem[{\citenamefont{Paessens and Sch\"utz}(2004)}]{paessens:jpa04}
\bibinfo{author}{\bibfnamefont{M.}~\bibnamefont{Paessens}} \bibnamefont{and}
  \bibinfo{author}{\bibfnamefont{G.~M.} \bibnamefont{Sch\"utz}},
  \bibinfo{journal}{J. Phys. A} \textbf{\bibinfo{volume}{37}},
  \bibinfo{pages}{4709} (\bibinfo{year}{2004}).

\bibitem[{\citenamefont{Suzuki}(1986)}]{suzuki:jpj86}
\bibinfo{author}{\bibfnamefont{M.}~\bibnamefont{Suzuki}}, \bibinfo{journal}{J.
  Phys. Soc. Jpn.} \textbf{\bibinfo{volume}{55}}, \bibinfo{pages}{4205}
  (\bibinfo{year}{1986}).

\bibitem[{\citenamefont{Howard and T\"auber}(1997)}]{howard:jpa97}
\bibinfo{author}{\bibfnamefont{M.~J.} \bibnamefont{Howard}} \bibnamefont{and}
  \bibinfo{author}{\bibfnamefont{U.~C.} \bibnamefont{T\"auber}},
  \bibinfo{journal}{J. Phys. A} \textbf{\bibinfo{volume}{30}},
  \bibinfo{pages}{7721} (\bibinfo{year}{1997}).

\bibitem[{\citenamefont{Jensen}(1993)}]{jensen:prl93}
\bibinfo{author}{\bibfnamefont{I.}~\bibnamefont{Jensen}},
  \bibinfo{journal}{Phys. Rev. Lett.} \textbf{\bibinfo{volume}{70}},
  \bibinfo{pages}{1465} (\bibinfo{year}{1993}).

\bibitem[{\citenamefont{Mu{\~n}oz et~al.}(1996)\citenamefont{Mu{\~n}oz,
  Grinstein, Dickman, and Livi}}]{munoz:prl96}
\bibinfo{author}{\bibfnamefont{M.~A.} \bibnamefont{Mu{\~n}oz}},
  \bibinfo{author}{\bibfnamefont{G.}~\bibnamefont{Grinstein}},
  \bibinfo{author}{\bibfnamefont{R.}~\bibnamefont{Dickman}}, \bibnamefont{and}
  \bibinfo{author}{\bibfnamefont{R.}~\bibnamefont{Livi}},
  \bibinfo{journal}{Phys. Rev. Lett.} \textbf{\bibinfo{volume}{76}},
  \bibinfo{pages}{451} (\bibinfo{year}{1996}).

\bibitem[{\citenamefont{Dickman}(1987)}]{dickman:pla87}
\bibinfo{author}{\bibfnamefont{R.}~\bibnamefont{Dickman}},
  \bibinfo{journal}{Phys. Lett. A} \textbf{\bibinfo{volume}{122}},
  \bibinfo{pages}{463} (\bibinfo{year}{1987}).

\bibitem[{\citenamefont{Marro and Dickman}(1999)}]{marro:99}
\bibinfo{author}{\bibfnamefont{J.}~\bibnamefont{Marro}} \bibnamefont{and}
  \bibinfo{author}{\bibfnamefont{R.}~\bibnamefont{Dickman}},
  \emph{\bibinfo{title}{Nonequilibrium Phase Transitions in Lattice Models}}
  (\bibinfo{publisher}{Cambridge University Press},
  \bibinfo{address}{Cambridge}, \bibinfo{year}{1999}).

\bibitem[{\citenamefont{Gutowitz et~al.}(1987)\citenamefont{Gutowitz, Victor,
  and Knight}}]{gutowitz:pd87}
\bibinfo{author}{\bibfnamefont{H.~A.} \bibnamefont{Gutowitz}},
  \bibinfo{author}{\bibfnamefont{J.~D.} \bibnamefont{Victor}},
  \bibnamefont{and} \bibinfo{author}{\bibfnamefont{B.~W.}
  \bibnamefont{Knight}}, \bibinfo{journal}{Physica D}
  \textbf{\bibinfo{volume}{28}}, \bibinfo{pages}{18} (\bibinfo{year}{1987}).

\bibitem[{\citenamefont{Szab\'o et~al.}(1990)\citenamefont{Szab\'o, Szolnoki,
  and Bod\'ocs}}]{szabo:pra91}
\bibinfo{author}{\bibfnamefont{G.}~\bibnamefont{Szab\'o}},
  \bibinfo{author}{\bibfnamefont{A.}~\bibnamefont{Szolnoki}}, \bibnamefont{and}
  \bibinfo{author}{\bibfnamefont{L.}~\bibnamefont{Bod\'ocs}},
  \bibinfo{journal}{Phys. Rev. A} \textbf{\bibinfo{volume}{44}},
  \bibinfo{pages}{6375} (\bibinfo{year}{1990}).

\bibitem[{\citenamefont{ben Avraham and K\"ohler}(1992)}]{benavraham:pra92}
\bibinfo{author}{\bibfnamefont{D.}~\bibnamefont{ben Avraham}} \bibnamefont{and}
  \bibinfo{author}{\bibfnamefont{J.}~\bibnamefont{K\"ohler}},
  \bibinfo{journal}{Phys. Rev. A} \textbf{\bibinfo{volume}{45}},
  \bibinfo{pages}{8358} (\bibinfo{year}{1992}).

\bibitem[{\citenamefont{Szab\'o and Borsos}(1994)}]{szabo:pre94}
\bibinfo{author}{\bibfnamefont{G.}~\bibnamefont{Szab\'o}} \bibnamefont{and}
  \bibinfo{author}{\bibfnamefont{I.}~\bibnamefont{Borsos}},
  \bibinfo{journal}{Phys. Rev. E} \textbf{\bibinfo{volume}{49}},
  \bibinfo{pages}{5900} (\bibinfo{year}{1994}).

\bibitem[{\citenamefont{Marques et~al.}(2002)\citenamefont{Marques, Santos, and
  Mendes}}]{marques:pre02}
\bibinfo{author}{\bibfnamefont{M.~C.} \bibnamefont{Marques}},
  \bibinfo{author}{\bibfnamefont{M.~A.} \bibnamefont{Santos}},
  \bibnamefont{and} \bibinfo{author}{\bibfnamefont{J.~F.~F.}
  \bibnamefont{Mendes}}, \bibinfo{journal}{Phys. Rev. E}
  \textbf{\bibinfo{volume}{65}}, \bibinfo{pages}{016111}
  (\bibinfo{year}{2002}).

\bibitem[{\citenamefont{Szolnoki}(2000)}]{szolnoki:pre00}
\bibinfo{author}{\bibfnamefont{A.}~\bibnamefont{Szolnoki}},
  \bibinfo{journal}{Phys. Rev. E} \textbf{\bibinfo{volume}{62}},
  \bibinfo{pages}{7466} (\bibinfo{year}{2000}).

\bibitem[{\citenamefont{Szolnoki}(2002)}]{szolnoki:pre02}
\bibinfo{author}{\bibfnamefont{A.}~\bibnamefont{Szolnoki}},
  \bibinfo{journal}{Phys. Rev. E} \textbf{\bibinfo{volume}{66}},
  \bibinfo{pages}{057102} (\bibinfo{year}{2002}).

\bibitem[{\citenamefont{Szab\'o and Toke}(1998)}]{szabo:pre98}
\bibinfo{author}{\bibfnamefont{G.}~\bibnamefont{Szab\'o}} \bibnamefont{and}
  \bibinfo{author}{\bibfnamefont{C.}~\bibnamefont{Toke}},
  \bibinfo{journal}{Phys. Rev. E} \textbf{\bibinfo{volume}{58}},
  \bibinfo{pages}{69} (\bibinfo{year}{1998}).

\bibitem[{\citenamefont{Dickman}(2002)}]{dickman:pre02b}
\bibinfo{author}{\bibfnamefont{R.}~\bibnamefont{Dickman}},
  \bibinfo{journal}{Phys. Rev. E} \textbf{\bibinfo{volume}{66}},
  \bibinfo{pages}{036122} (\bibinfo{year}{2002}).

\bibitem[{\citenamefont{Park and Park}(2004)}]{park:cm04}
\bibinfo{author}{\bibfnamefont{S.-C.} \bibnamefont{Park}} \bibnamefont{and}
  \bibinfo{author}{\bibfnamefont{H.}~\bibnamefont{Park}},
  \bibinfo{journal}{condmat/0409115}  (\bibinfo{year}{2004}).

\bibitem[{\citenamefont{da~Silva and Dickman}(1999)}]{kamphorst:pre99}
\bibinfo{author}{\bibfnamefont{J.~K.~L.} \bibnamefont{da~Silva}}
  \bibnamefont{and} \bibinfo{author}{\bibfnamefont{R.}~\bibnamefont{Dickman}},
  \bibinfo{journal}{Phys. Rev. E} \textbf{\bibinfo{volume}{60}},
  \bibinfo{pages}{5126} (\bibinfo{year}{1999}).

\bibitem[{\citenamefont{Zhong et~al.}(2003)\citenamefont{Zhong, ben Avraham,
  and Mu{\~n}oz}}]{zhong:epjb03}
\bibinfo{author}{\bibfnamefont{D.}~\bibnamefont{Zhong}},
  \bibinfo{author}{\bibfnamefont{D.}~\bibnamefont{ben Avraham}},
  \bibnamefont{and} \bibinfo{author}{\bibfnamefont{M.~A.}
  \bibnamefont{Mu{\~n}oz}}, \bibinfo{journal}{Europ. Phys. J. B}
  \textbf{\bibinfo{volume}{35}}, \bibinfo{pages}{505} (\bibinfo{year}{2003}).

\end{thebibliography}
\end{document}